\documentclass[runningheads]{llncs}
\usepackage{graphicx}
\usepackage{comment}
\usepackage{url}
\usepackage{subcaption}
\usepackage{calc}
\usepackage{rotating}
\usepackage{booktabs}
\usepackage{hyperref}
\begin{document}
\title{Representing Semantified Biological Assays in the Open Research Knowledge Graph\thanks{Supported by TIB Leibniz Information Centre for Science and Technology, the EU H2020 ERC project ScienceGRaph (GA ID: 819536) and the ITN PERICO (GA ID: 812968).}}
\titlerunning{Semantified Bioassays in the Open Research Knowledge Graph}
%
\author{Marco Anteghini\inst{1,2}\orcidID{0000-0003-2794-3853} \and Jennifer D'Souza\inst{3}\orcidID{0000-0002-6616-9509}\and Vitor A.P. Martins dos Santos\inst{1,2}\orcidID{0000-0002-2352-9017} \and S\"oren Auer\inst{3}\orcidID{0000-0002-0698-2864}}
\authorrunning{Anteghini et al.}
%
\institute{Lifeglimmer GmbH, Markelstr. 38, 12163 Berlin, Germany
\and
Wageningen University \& Research, Laboratory of Systems \& Synthetic Biology, Stippeneng 4, 6708 WE, Wageningen, The Netherlands 
\email{\{anteghini,vds\}@lifeglimmer.com} \\
\and
TIB Leibniz Information Centre for Science and Technology, Hannover, Germany \\
\email{\{jennifer.dsouza,soeren.auer\}@tib.eu}}
\maketitle              
\begin{abstract} In the biotechnology and biomedical domains, recent text mining efforts advocate for machine-interpretable, and preferably, semantified, documentation formats of laboratory processes. This includes wet-lab protocols, (in)organic materials synthesis reactions, genetic manipulations and procedures for faster computer-mediated analysis and predictions. Herein, we present our work on \textit{the representation of semantified bioassays in the Open Research Knowledge Graph (ORKG)}. In particular, we describe a semantification system work-in-progress to generate, automatically and quickly, the critical semantified bioassay data mass needed to foster a consistent user audience to adopt the ORKG for recording their bioassays and facilitate the organisation of research, according to FAIR principles.

\end{abstract}
\section{Introduction}

More and more scholarly digital library initiatives aim at fostering the digitalization of traditional document-based scholarly articles~\cite{aryani2018research,auer_soren_2018,baas2020scopus,birkle2020web,fricke2018semantic,hendricks2020crossref,manghi_paolo_2019_3516918,wang2020microsoft}. This means structuring and organizing, in a fine-grained manner, knowledge elements from previously unstructured scholarly articles in a Knowledge Graph. These efforts are analogous to the digital transformation seen in recent years in other information-rich publishing and communication services, e.g., e-commerce product catalogs instead of mailorder catalogs, or online map services instead of printed street maps. For these services, the traditional document-based publication was not just digitized (by making digitized PDFs of the analog artifacts available) but has seen a comprehensively transformative digitalization.

Of available scholarly knowledge digitalization avenues~\cite{aryani2018research,auer_soren_2018,baas2020scopus,birkle2020web,fricke2018semantic,hendricks2020crossref,manghi_paolo_2019_3516918}, we highlight the Open Research Knowledge Graph (ORKG)~\cite{jaradeh2019open}. It is a next-generation digital library (DL) that focuses on ingesting information in scholarly articles as machine-actionable knowledge graphs (KG). In it, an article is represented with both (bibliographic) metadata and semantic descriptions (as subject-predicate-object triples) of its \textit{contributions}. ORKG has a number of advantages as: 1) it enables flexible semantic content modeling (i.e., ontologized or not, depending on the user or domain); 2) it semantifies \textit{contributions} at various levels of granularity from shallow to fine-grained; and 3) it publishes persistent KG links per article contribution that it contains. For further technical details about the platform, we refer the reader to the introductory paper~\cite{jaradeh2019open}.

The ORKG DL aims to integrate and interlink contributions' KGs for Science at large, i.e. multidisciplinarily. Thus far, ongoing efforts are in place for integrating scholarly contributions from at least two disciplines, viz. Math~\cite{math} (e.g., \url{https://www.orkg.org/orkg/paper/R12192}) and the Natural Language Processing subdomain in AI~\cite{nlpcontributions} (e.g., \url{https://www.orkg.org/orkg/paper/R44253}). Moreover, the ORKG also has a separate feature to automatically import individual articles' contributions data found tabulated in survey articles~\cite{oelen}. E.g., an ORKG object for Earth Science articles' contributions surveyed: \url{https://www.orkg.org/orkg/comparison/R38484}. Since surveys are written in most disciplines, this latter feature directly targets the ORKG aim; however, its sole limitation is that it is restricted only to those papers that have been surveyed. On the other hand, with the per-domain semantification models, articles not surveyed can be also modeled in the ORKG.

In this paper, we describe our ongoing work in extending the ORKG to integrate biological assays from the Biochemistry discipline. For bioassays, a semantification model already exists as the BioAssay Ontology (BAO)~\cite{bao}. However, we need to design a pragmatic workflow for integrating bioassays semantified by the BAO in the ORKG DL. To this end, we discuss the manual and automatic process of integrating such semantified data in the ORKG DL. Furthermore, we show how these semantified data integrated in the ORKG is amenable to advanced computational processing support for the researcher. 

With the volume of research burgeoning~\cite{johnson2018stm}, adopting a finer-grained semantification as KG for scholarly content representation is compelling. Better semantification means better machine actionability, which in turn means innumerable possibilities of advanced computational functions on scholarly content. One function especially poignant in this era of the publications deluge~\cite{jinha2010article}, is computational support to alleviate the manual information ingestion cognitive burden. This is precisely the computational support showcase we depict from the ORKG DL over our integrated bioassay KGs, consequently highlighting the benefits of digitalizing bioassays and of the ORKG DL platform.

\section{Our Work-In-Progress Aims and Motivations}

\begin{quote}
\textit{Allowing practitioners to easily search for similar bioassays as well as compare these semantically structured bioassays on their key properties.}
\end{quote}

\paragraph{Why integrate bioassays in a knowledge graph?} 
Until their recent semantification in an expert-annotated dataset of 983 bioassays~\cite{clark2014fast,schurer2011bioassay,vempati2012formalization} based on the BAO~\cite{bao}, bioassays were published in the form of plain text. Integrating their semantified counterpart in a KG facilitates their advanced computational processing. Consider that key assay concepts related to biological screening, including Perturbagen, Participants, Meta Target and Detection Technology, will be machine-actionable. This widens the potential for relational enrichment and interlinking when integrated with machine-interpretable formats of wet lab protocols and inorganic materials synthesis reactions and procedures~\cite{chemrecipes,labprotocols,kuniyoshi2020annotating,mysore2019materials}. 
Furthermore, in this era of neural-based ML technologies, KG-based word embeddings foster new inferential discovery mechanisms given that they encode high-dimensional semantic spaces~\cite{bianchi2020knowledge} with bioassay KGs so far untested for.   

\paragraph{Why the ORKG DL}\cite{auer_soren_2018}? The core of the setup of knowledge-based digitalized information flows is the distributed, decentralized, collaborative creation and evolution of information models. Moreover, vocabularies, ontologies, and knowledge graphs to establish a common understanding of the data between the various stakeholders. And, importantly, the integration of these technologies into the infrastructure and processes of search and knowledge exchange toward a research library of the future. The ORKG DL is such a solution. Implemented within TIB, as a central library and information centre for science and technology, it also promises development longevity: the Leibniz Association institutional networks presents a critical mass of application domains and users to enhance the infrastructure and continuously integrate new knowledge disciplines.




With these considerations in place, the work described in the subsequent sections is being carried forth. Next, we describe our approach in the context of two main research questions.


\begin{figure*}[!tb]
  \centering
  \includegraphics[width=\textwidth]{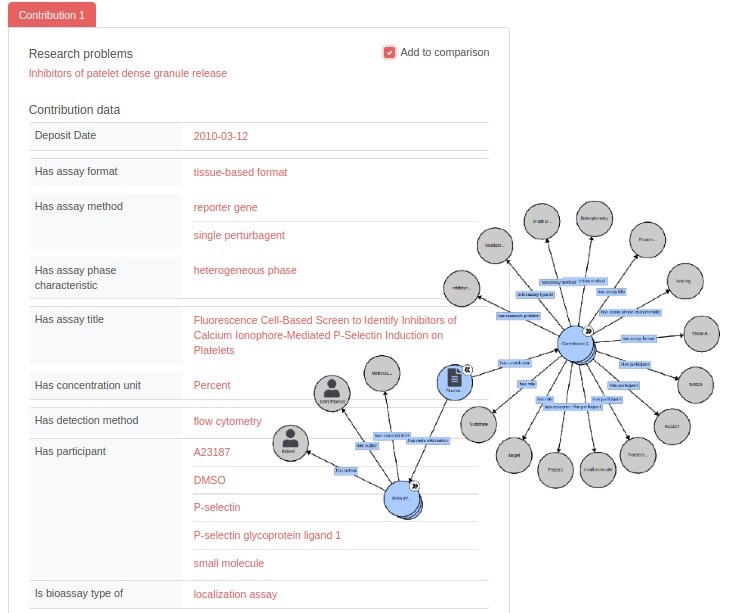}
\caption{An ORKG representation of a semantified Bioassay with an overlayed graph view of the assay. Accessible at: \href{https://www.orkg.org/orkg/paper/R48178}{https://www.orkg.org/orkg/paper/R48178}}
\label{fig:fig1}
\end{figure*}

\section{Approach: Digitalization of Biological Assays}


\textbf{RQ1:} \textit{What are steps for manually digitalizing a Bioassay in the ORKG?} The digitalization is based on the prior requirement that text-based bioassays are semantified based on the BioAssay Ontology (BAO)~\cite{bao}. This is the manual aspect of the digitalization process involving domain experts or the assay authors themselves. In Figure~\ref{fig:fig1}, we show an example of a manually pre-semantified bioassay integrated in ORKG. This bioassay was semantified on eight properties based on the BAO. It was drawn from an expert-annotated set of 983 bioassays~\cite{schurer2011bioassay,vempati2012formalization}. In terms of salient features, the bioassays in this dataset have 53 triple semantic statements on average with a minimum of 5 and a maximum of 92 statements; there are 42 different types of bioassays (e.g., luciferase reporter gene assay, protein-protein interaction assay---see in appendix the full list); and there are 11 assay formats (e.g., cell-based, biochemical). Thus, the manual semantification task complexity can be viewed as 53 modeling decisions.

In gist, the manual digitalizaton of a bioassay in the ORKG includes: 1) \textit{a BAO-based semantification step}: forming subject-predicate-object triples of the bioassay text content based on the BAO. E.g., for the assay in Fig.~\ref{fig:fig1}, a few of its semantic triples are: (Contribution, Has assay format, tissue-based format), (Contribution, Has assay method, reporter gene), among others. And as a recommended step, 2) associating each ontologized resource (i.e., a subject, a predicate, an object) with a URI as its defining class in the original ontology, which for bioassays is the BAO.

Having just described the manual digitalization workflow, we next present our hybrid workflow that is currently in development. In this, we decide to incorporate automated semantification which levies pragmatic considerations in the digitalization of bioassays in the ORKG. Relatedly, there is an existing hybrid system~\cite{clark2014fast} for semantifying bioassays involving machine learning and expert interaction which inspires our work. Nonetheless, we differ. While their learning-based component relies heavily on explicitly encoded syntactic features of the text, ours relies on neural networks based on the current state-of-the-art transformer models~\cite{vaswani2017attention} trained on millions of scientific articles~\cite{beltagy2019scibert}. Such systems by encoding high-dimensional semantic spaces of the underlying text, obviate the need to make explicit considerations for features of the text. Moreover, they significantly outperform systems designed based on explicit features~\cite{devlin2019bert}---with due credit to the system by Clark et al.~\cite{clark2014fast} designed prior to the onset of this revolutionary technology. Next, our hybrid workflow is designed toward a practical end---to be integrated in the ORKG DL which has a predominant focus on the digitalization of scholarly knowledge content multidisciplinarily, thus setting it apart from any existing DL. 

\textbf{RQ2:} \textit{What are the modules needed in the hybrid digitalization of Bioassays in the ORKG?} Essentially, given a new bioassay text input, we are implementing two modules in a two-step workflow as follows: 1) an automated semantifier; and 2) a human-in-the-loop curation of the predicted labels either by the assay author or a dedicated curator. Unlike the manual workflow, this presents a much easier and less time-intensive task for the human. They would be merely selecting the correctly predicted triples, deleting the incorrect ones, or defining new ones as needed. Assuming a well-trained machine learning module, the latter two steps may be entirely omitted. Toward this hybrid workflow, as work in progress, the automated semantifier is in development, and we are also implementing extensions in the ORKG infrastructure to include additional front-end views as assay curation interfaces.

\section{Solving the Cognitive Information Ingestion Hurdle: Comparison Surveys across KG-based Bioassays}

\begin{figure*}[!tb]
  \centering
  \includegraphics[width=\textwidth]{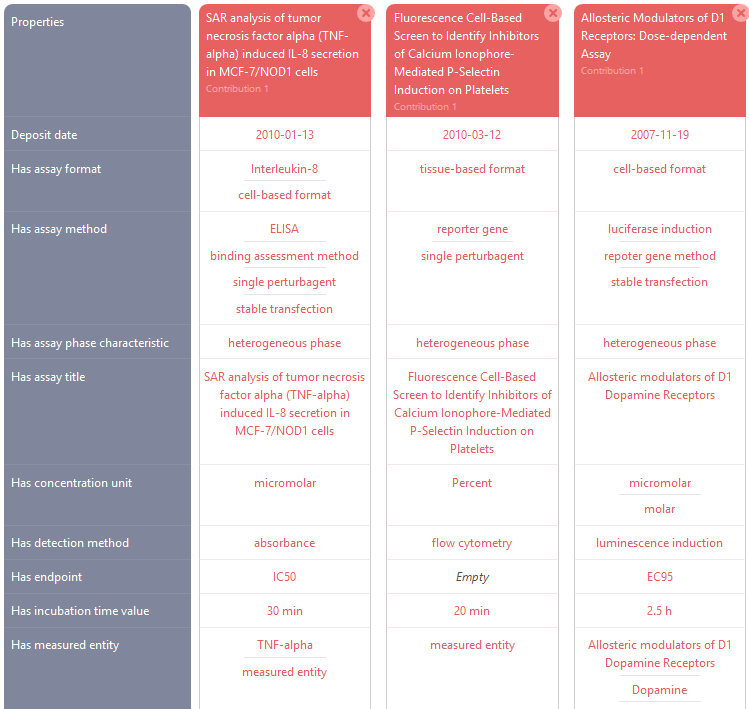}
\caption{Comparisons of semantified bioassays in the ORKG digital library. Online \footnotesize{\url{https://www.orkg.org/orkg/comparison?contributions=R48195,R48179,R48147}}}
\label{fig:fig-compare}
\end{figure*}

\textit{Premise: We need an information processing tool that can be used by biomedical practitioners to quickly comprehend bioassays' key properties.} 

The ORKG DL has a computational feature to generate and publish surveys in the form of a tabulated comparisons of the KG nodes~\cite{oelen}. To demonstrate this feature, we manually entered the data of three semantified bioassays in the ORKG DL. Applying then the ORKG survey feature on the three assays aggregates their semantified graph nodes in tabulated comparisons across the assays. This is depicted in Figure~\ref{fig:fig-compare}. With such structured computations enabled, we have a novel approach to uncovering and presenting information relying on aggregated scholarly knowledge. 
The computation shown in Fig.~\ref{fig:fig-compare} aligns closely with the notion of the traditional survey articles, except it is fully automated and operates on machine-actionable knowledge elements.
The BAO-semantified assays are compared side-by-side on their graph nodes. Thus, tracking the progress on bioassays, can be eased from a task of several days to a few minutes. 


\section{Conclusion}

Thus in this paper, we outlined a vision in two separate workflows for integrating bioassay knowledge in the ORKG DL and our ongoing work to this end. The implications of bioassay structured and machine-actionable knowledge are broad. To mention just one in the particular context of the current Covid-19 pandemic: The discovery of cures for diseases can be greatly expedited if scientists are given intelligent information access tools, and our work toward automatically semantifying bioassays are a step in this direction.

To this end, the workflows prescribed in this work offer the possibilities to chose between a manual or a semi-automatic strategy for bioassays' semantification within a real-world digital library. 

We would like to invite interested researchers to collaborate with us on the following topics: 1) generating a large dataset of semantically structured bioassays; 2) user evaluation of our semi-automated system for semantically structuring bioassay data.

We deem this as a starting point for a discussion in the community ultimately leading to more clearly defined technical requirements, and a roadmap for fulfilling the potential of the ORKG as a next-generation digital library for fine-grained semantified access to scholarly content.

%
%
%
\bibliographystyle{splncs04}
\bibliography{mybib}
\clearpage
\appendix

\section{Bioassay types}
\begin{table}[]
    \centering
    \begin{tabular}{p{6cm}p{6cm}}\toprule
\multicolumn{2}{c}{\textbf{Bioassay types}}\\
\\\midrule
protein-protein interaction & hydrolase activity \\
kinase activity & protein-small molecule interaction \\
viability & beta lactamase reporter gene \\
cytochrome P450 enzyme activity & luciferase enzyme activity\\
luciferase reporter gene & oxidoreductase activity \\
protein unfolding & chaperone activity \\
lyase activity & transporter \\
plasma membrane potential & dye redistribution\\
calcium redistribution & apoptosis \\
beta lactamase reporter gene & beta galactosidase reporter gene\\
phosphatase activity& cAMP redistribution\\
IP1 redistribution & cell morphology \\
phosphorylation & transferase activity  \\
isomerase activity  & protein redistribution \\
radioligand binding & signal transduction \\
ion channel & platelet activation\\
fluorescent protein reporter gene& protein-DNA interaction\\
protease activity & cell permeability \\
protein stability & protein-turnover \\
localization & organism behavior\\
cytotoxicity& cell growth\\\bottomrule
\\
    \end{tabular}
    \caption{List of the different bioassay types present in our dataset}
    \label{tab:my_label}
\end{table}

\section{Preliminary Results of Automated Semantification: SciBERT-based Bioassay Semantifier}

\begin{figure*}[!b]
  \centering
  \includegraphics[width=0.98\textwidth,height=16cm]{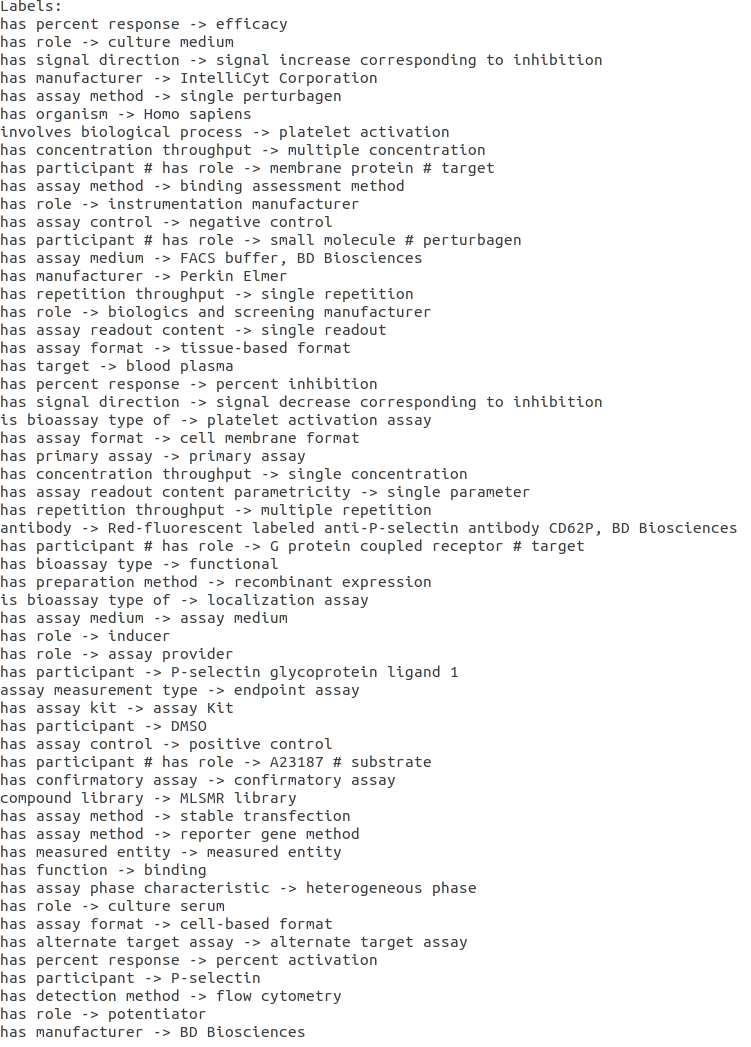}
\caption{Automatically semantified bioassay (human-annotated reference in Fig.~\ref{fig:fig1})}
\label{fig:fig3}
\end{figure*}

The semantic statements depicted in Figure~\ref{fig:fig3} were automatically generated from SciBERT-based~\cite{beltagy2019scibert} neural semantification system. These predictions were made for the same bioassay text depicted in Figure~\ref{fig:fig1}. Comparing the automatically generated one against the reference, we see that almost all the manually curated labels are correctly predicted. Among 16 manually curated labels, excluding those we omit in our training procedure (e.g., has title, PubChem AID, Deposit Date, has incubation time value, has concentration unit), the model accurately predicts 12 statements, while the remaining were deemed by a domain-specialist as valid additional candidates to incorporate in the reference set (e.g., has significant direction, has concentration throughput). 

\end{document}